\documentclass[twocolumn,showpacs,preprintnumbers,amsmath,amssymb]{revtex4}

\newcommand{\bib}{\bibitem}
\newcommand{\bea}{\begin{eqnarray}}
\newcommand{\eea}{\end{eqnarray}}
\newcommand{\beq}{\begin{equation}}
\newcommand{\eeq}{\end{equation}}
\newcommand{\non}{\nonumber}
\newcommand{\noi}{\noindent}
\newcommand{\da}{\dagger}
\newcommand{\al}{\alpha}
\newcommand{\be}{\beta}
\newcommand{\ga}{\gamma}

\newcommand{\De}{\Delta}

\newcommand{\ka}{\kappa}
\newcommand{\om}{\omega}

\newcommand{\ct}{\cite}

\usepackage{graphicx,epsfig} 
\usepackage{dcolumn} 
\usepackage{bm} 

\begin{document}

\title{Nonadiabatic charge pumping in a one-dimensional system of
noninteracting electrons by an oscillating potential}
\author{Amit Agarwal and Diptiman Sen}
\affiliation{Center for High Energy Physics, Indian Institute of Science,
Bangalore 560012, India}
\date{\today}

\begin{abstract}
Using a tight-binding model, we study one-parameter charge pumping in a 
one-dimensional system of non-interacting electrons. An oscillating potential
is applied at one site while a static potential is applied in a different 
region. Using Floquet scattering theory, we calculate the current up to second
order in the oscillation amplitude and exactly in the oscillation frequency. 
For low frequency, the charge pumped per cycle is proportional to the 
frequency and therefore vanishes in the adiabatic limit. If the static 
potential has a bound state, we find that such a state has a significant 
effect on the pumped charge if the oscillating potential can excite the 
bound state into the continuum states or vice versa. Finally, we use the 
equation of motion for the density matrix to numerically compute the pumped 
current for any value of the amplitude and frequency. The numerical results 
confirm the unusual effect of a bound state. 
\end{abstract}

\maketitle

\section{Introduction}

The phenomenon of charge pumping and rectification by time-dependent 
potentials applied to certain points in a system has been extensively 
studied both theoretically [1-31] and experimentally 
\ct{switkes,taly1,cunningham,taly2,leek,dicarlo,kaestner}. The idea of charge
pumping is that periodically oscillating potentials can transfer a net charge
per cycle between two leads which are at the same chemical potential. For the 
case of non-interacting electrons, theoretical studies of charge pumping have 
used adiabatic scattering theory \ct{avron,entin1,entin2,hwang}, Floquet 
scattering theory \ct{moskalets,kim}, variations of the non-equilibrium Green 
function formalism \ct{wang,arrachea1,torres}, and the equation of motion 
approach \ct{agarwal1}. The case of interacting electrons has also been 
studied, using a renormalization group method for weak interactions \ct{das},
and the method of bosonization for arbitrary interactions [40-50]. 

Apart from a few papers \ct{arrachea1,torres,kaestner}, the earlier 
studies of charge pumping have generally studied systems in which oscillating
potentials are applied to two or more sites. In such cases, it is known that 
if the oscillation frequency $\om$ is small, the dc part of the pumped current
is proportional to $\om$; the charge pumped per cycle (with time period
$2\pi /\om$) therefore has a finite value in the adiabatic limit $\om \to 0$.
However, an oscillating potential applied to a single site can also pump
charge provided that the system has no left-right symmetry, as has been
emphasized in Refs. \ct{arrachea1,torres}. This can happen
if, for instance, there is a static potential at another site. In these cases,
however, the dc part of the pumped current is proportional to $\om^2$, and the
charge pumped per cycle is proportional to $\om$ if $\om$ is small. In this 
paper, we will study 
such a system in detail using both analytical and numerical 
methods. Our analysis will not be restricted to small values of $\om$. If 
$\om$ is large enough, we discover that a bound state (defined as a state whose
energy lies outside the continuum band of the tight-binding model that we will 
consider) can have a significant effect on the pumped charge. Namely, although
the bound state has a localized wave function and therefore cannot contribute
to the current at the same energy, the electrons can be scattered from the 
bound state to the continuum states (lying within the band) or vice versa, 
and this can contribute to the current flow. This phenomenon does not seem
to have received much attention in the existing literature.

The plan of the paper is as follows. In Sec. II, we will use Floquet 
scattering theory to obtain an expression for the pumped current produced by 
a single harmonically oscillating potential, when there is
a static potential present at some other point in the system. Our analysis
will be exact in the scattering matrix arising from the static potential and 
in the oscillation frequency, but it will be perturbative in the amplitude 
of the oscillations. The effect of a bound state produced by the static 
potential will also be considered using the same formalism. In Sec. III, we 
will use the equation of motion to numerically compute the pumped charge for 
the same model; this method can be used for any value of the oscillation 
amplitude. Our numerical results will confirm the unusual effect that a 
bound state can have on the pumped charge, as well as the difficulty which
the equation of motion method faces in dealing with a bound state \ct{dhar}.
We will summarize our results in Sec. IV. 

\section{Floquet scattering theory}

Let us consider a one-dimensional system consisting of two semi-infinite 
leads $a=L,R$ (denoting left and right) and a finite region called the wire 
which lies between the two. We will model the three regions together by a 
lattice model of spinless electrons governed by a one-channel tight-binding 
Hamiltonian with the same hopping amplitude $-\ga$ on all bonds, namely,
\bea {\hat H}_0 ~=~ -\ga ~\sum_{n=-\infty}^{\infty} (~ c_{n+1}^\da c_n ~+~ 
c_n^\da c_{n+1} ) ~+~ a ~c_0^\da c_0 , \label{h0} \eea
where we have added a time-independent potential with strength $a$ at the 
site $n=0$. The dispersion of the electrons in the leads is $E_k = -2\ga 
\cos k$, where $k$ lies in the range $[-\pi ,\pi]$. (We are setting the 
Planck constant $\hbar$ and the lattice spacing equal to unity). The two leads
are assumed to be at the same chemical potential $\mu$ and temperature $T$. 

The static potential at the site $n=0$ causes scattering of electrons 
incident from the left or right lead. If an incident electron has wave 
number $k$, the effect of the potential is described by a scattering matrix
\bea S (k) &=& \left( \begin{array}{cc} r_L & t_L \\
t_R & r_R \end{array} \right), \non \\
{\rm where} ~~~r_R (k) &=& r_L (k) ~=~ -~ \frac{ia}{2 \ga \sin k ~+~ ia}, 
\non \\ 
t_R (k) &=& t_L (k) ~=~ \frac{2 \ga \sin k}{2 \ga \sin k ~+~ ia},
\label{smat} \eea
and $r_{L(R)}$ and $t_{R(L)}$ denote the reflection and transmission 
amplitudes for an electron coming from the left (right) respectively.

It turns out that for any value of the parameter $a \ne 0$, there is a bound 
state. For $a > 0$, the bound state energy $E_B = \sqrt{4 \ga^2 + a^2}$ lies 
above the continuum, and the normalized bound state wave function is 
\beq \psi_B (n) ~=~ (-1)^n ~\sqrt{\tanh \ka} ~e^{-\ka |n|} ~~~{\rm for ~all}~~
n, \label{bs1} \eeq
where $\ka = \sinh^{-1} (a /2\ga)$. For $a < 0$, the bound state energy $E_B 
= - \sqrt{4 \ga^2 + a^2}$ lies below the continuum, and the corresponding wave
function is
\beq \psi_B (n) ~=~ \sqrt{\tanh \ka} ~e^{-\ka |n|} ~~~{\rm for ~all}~~ n, 
\label{bs2} \eeq
where $\ka = \sinh^{-1} (-a /2\ga)$.

We now apply an oscillating potential at the site $n=r$ of the model
described in Eq. (\ref{h0}), where we assume that $r \ge 1$. This part of 
the Hamiltonian is given by
\beq {\hat V} (t) ~=~ b ~\cos (\om t) ~c_r^\da c_r. \label{pot} \eeq
Then Floquet scattering theory works as follows \ct{moskalets,kim}. 
Incoming electrons of energy $E_0$ gain or lose energy in quanta of $\om$ on
interacting with the oscillating potential. Hence, the outgoing states are
characterized by energies $E_p = E_0 + p \om$, where $p=0,\pm 1,\pm2, \cdots$;
the energies with $p \ne 0$ are called the Floquet side bands.
The effect of the oscillating potential can be described by a Floquet
scattering matrix $S_{\al \be} (E_p, E_0)$, which is the amplitude for
an electron with energy $E_0$ entering through lead $\be$ to leave with
energy $E_p$ through lead $\al$. In the leads, the propagating modes have 
energies lying within the continuum band $[-2\ga ,2 \ga]$; only these modes 
can directly contribute to charge pumping. States with energies lying outside 
the continuum band have wave functions which decay exponentially into the 
leads and hence do not directly contribute to charge transfer. The wave 
function of an electron coming from the left lead with an energy $E_0$
and wave number $k_0$ (with $E_0 = - 2 \ga \cos k_0$) is given by
\beq \psi (n) ~=~ e^{i(k_0n - E_0 t)} ~+~ \sum_p ~r_{L,p} ~e^{i(-k_p n - E_p 
t)}, \eeq
at a site $n$ far to the left of the scattering region, and
\beq \psi (n) ~=~ \sum_p ~t_{R,p} ~e^{i(k_p n - E_p t)}, \eeq
far to the right of the scattering region, where $E_p = - 2 \ga \cos k_p$, and
the sums over $p$ run over values such that $E_p$ lies within the continuum
band of the leads. The quantities $r_{L,p}$ and $t_{R,p}$ denote reflection 
and transmission amplitudes in the different side bands; they respectively 
denote the elements $S_{LL} (E_p, E_0)$ and $S_{RL} (E_p, E_0)$ of the Floquet
scattering matrix, where $L$ and $R$ denote the left and right leads. 
Similarly, the wave function of an electron coming from the right lead with 
an energy $E_0$ and wave number $k_0$ is given by
\beq \psi (n) ~=~ e^{i(-k_0n - E_0 t)} ~+~ \sum_p ~r_{R,p} ~e^{i(k_p n - E_p 
t)}, \eeq
far to the right of the scattering region, and
\beq \psi (n) ~=~ \sum_p ~t_{L,p} ~e^{i(-k_p n - E_p t)}, \eeq
far to the left of the scattering region. Due to unitarity, we have the 
relations
\bea \sum_p ~\frac{v_p}{v_0} ~[~ |r_{L,p}|^2 ~+~ |t_{R,p}|^2 ~] &=& 1, \non \\
{\rm and} ~~~\sum_p ~\frac{v_p}{v_0} ~[~ |r_{R,p}|^2 ~+~|t_{L,p}|^2 ~] &=& 1, 
\label{unit} \eea
where $v_p = 2 \ga \sin k_p$ is the velocity in the $p$-th side band.

The reflection and transmission amplitudes are found by writing down the wave 
functions in the scattering region, and matching coefficients of terms having 
the same time dependence ($e^{\pm i E_p t}$) in the Schr\"odinger equation at 
different sites. If the oscillating potentials are weak, the reflection and 
transmission amplitudes decrease rapidly as $|p|$ increases; at first order 
in the potentials, only $p = \pm 1$ contribute. The dc part of the current 
in, say, the right lead is then given by
\bea I_R & & = ~q ~\int_{-2\ga}^{2\ga - \om} ~\frac{dE_0}{2\pi} ~
\frac{v_1}{v_0} ~(|t_{R,1}|^2 + |r_{R,1}|^2 ) \non \\
& & ~~~~~~~~~~~~~~~~~~~~ \times ~\{ f(E_0) - f(E_1) \} \non \\
& & ~+ q ~\int_{-2\ga + \om}^{2\ga} ~\frac{dE_0}{2\pi} ~\frac{v_{-1}}{v_0}~
(|t_{R,-1}|^2 + |r_{R,-1}|^2 ) \non \\
& & ~~~~~~~~~~~~~~~~~~~~ \times ~\{ f(E_0) - f(E_{-1} \} ~], \label{ir1} \eea
where $f (E) ~=~ 1/[e^{(E - \mu)/k_B T} + 1]$ is the Fermi function, 
and $q$ is the charge of an electron. The upper
limit is $2 \ga - \om$ in the first integral in Eq. (\ref{ir1}) because if 
$E_0 > 2 \ga - \om$, $E_1$ will lie above the continuum band and will therefore
not contribute to the current. Similarly, the lower limit is $- 2 \ga + \om$ 
in the second integral in Eq. (\ref{ir1}) because if $E_0 < - 2 \ga + \om$, 
$E_{-1}$ will lie below the continuum band and will not contribute to the 
current. If $\om > 4 \ga$, the integrals will contribute nothing since all 
the side bands lie outside the continuum band; the pumped current will 
therefore vanish.

For the case of a single static potential and a single oscillating potential
described by Eqs. (\ref{h0}) and (\ref{pot}), we find that
\bea r_{R,1} &=& - ~\frac{ib}{4 \ga \sin k_1} ~e^{-i(k_0 + k_1) r} ~
[~ 1 ~+~ r_R (k_0) ~e^{i2k_0 r} ~] \non \\
& & \times ~~[~ 1 ~+~ r_R (k_1) ~e^{i2k_1r} ~], \non \\
t_{R,1} &=& - ~\frac{ib}{4 \ga \sin k_1} ~t_R (k_0) ~e^{i(k_0 - k_1)r} \non \\
& & \times ~~[~ 1 ~+~ r_R (k_1) ~e^{i2k_1r} ~], \label{rt} \eea
where the functions $r_R$ and $t_R$ are given in Eq. (\ref{smat}). The 
expressions for $r_{R,-1}$ and $t_{R,-1}$ can be obtained from Eq. (\ref{rt})
by replacing $k_1$ by $k_{-1}$. By substituting all this in Eq. (\ref{ir1}),
we obtain an expression for $I_R$ which is valid up to second order in the
dimensionless quantity $b/(2\ga \sin k_p)$.

In the limit $\om \to 0$, we have $f(E_{\pm 1}) - f(E_0) = \pm ~\om df(E_0) /
dE_0$. By shifting $E_0 \to E_0 + \om$ in the second integral in Eq.
(\ref{ir1}), one can show that $I_R$ goes as $\om^2$ for $\om \to 0$. In
particular, for zero temperature, $df(E_0) /dE_0 = - \delta (E_0 - \mu)$,
and we get 
\beq I_R ~=~ \frac{q\om^2 b^2 |t_R (k_F)|^4}{64 \pi \ga^3 \sin^3 k_F} \left(
\frac{d}{dk} ~\frac{|1 ~+~ r_R (k) ~e^{i2kr}|^2}{|t_R (k)|^2} \right)_{k=k_F},
\label{irzero} \eeq
where $k_F$ is the Fermi wave number given by $\mu = - 2 \ga \cos k_F$. Note 
that Eq. (\ref{irzero}) vanishes if $r=0$ since $1+r_R (k)= t_R (k)$; thus a 
non-zero value of $r$, i.e.,
a left-right spatial asymmetry, is necessary to have charge pumping. The 
expression in (\ref{irzero}) is in contrast to the case in which there 
are two or more oscillating potentials at different locations in which case 
the pumped current generally goes as $\om$ for $\om \to 0$. 

Equation (\ref{irzero}) can be generalized to the case where the time-dependent
potential at a given site has a finite number of oscillating terms each of 
which has a small amplitude and low frequency, namely, if $b \cos (\om t)$ in 
Eq. (\ref{pot}) is replaced by $\sum_i b_i \cos (\om_i t + \phi_i)$. To second
order in $b_i$ and $\om_i$, we find that the dc part of the pumped current at 
zero temperature can be obtained by replacing the factor $\om^2 b^2$ in Eq. 
(\ref{irzero}) by $\sum_i \om_i^2 b_i^2$.

We have so far discussed the contribution to the pumped current from the
scattering states only. It turns out that a bound state can also contribute
to the pumped current. Suppose that the static potential gives rise to a 
bound state with energy $E_B < - 2 \ga$. If the oscillation frequency is large
enough that $E_1 = E_B + \om$ lies within the continuum band $[-2\ga , 2\ga]$,
then the bound state will contribute to the current. Using
Floquet theory up to first order in $b$, we find that the wave function is 
given by $\al_R e^{i(k_1 n - E_1 t)}$ and $\al_L e^{i(-k_1 n - E_1 t)}$ far
to the right and left of the scattering region respectively, where
\bea \al_R &=& - ~\frac{ib}{4 \ga \sin k_1} ~e^{-ik_1 r} ~\psi_B (r) \non \\
& & \times ~~[~ 1 ~+~ r_R (k_1) ~e^{i2k_1r} ~], \non \\
\al_L &=& - ~\frac{ib}{4 \ga \sin k_1} ~e^{ik_1 r} ~\psi_B (r) ~t_L (k_1), 
\label{alr} \eea
and the function $\psi_B$ is given in Eq. (\ref{bs2}). The dc part of the
current pumped to the right is then given by 
\bea I_{RB} &=& q ~2 \ga \sin k_1 ~[~ |\al_R|^2 ~-~ |\al_L|^2 ~] ~\{ f(E_B) -
f(E_1) \} , \non \\
& & \label{irb} \eea
where $2 \ga \sin k_1$ is the electron velocity. Equation (\ref{irb}) has
to be added to Eq. (\ref{ir1}) in order to obtain the total pumped current.
At zero temperature, $I_{RB}$ is non-zero only if $E_B + \om > \mu >
E_B$, and it is then independent of $\mu$.

Similar considerations hold if there is a positive energy bound state with 
$E_B > 2 \gamma$ and $E_{-1} = E_B - \om$ lies within the continuum band.
Such a bound state will then contribute to the pumped current. One can
compute this contribution by applying a particle-hole transformation 
to the calculation given above for a negative energy bound state. Under
the transformation $c_n \to (-1)^n c_n$, the hopping term in Eq. (\ref{h0})
remains the same but $c_n^\da c_n$ changes sign. This is equivalent to 
changing $a \to -a$ in Eq. (\ref{h0}), and the chemical potential $\mu \to 
- \mu$. Thus the filling fraction $f = k_F /\pi$ changes as $f \to 1 - f$, 
and the current changes as $I_R \to - I_R$; the latter can be seen directly 
from the form of the current operator given in Eq. (\ref{jnt}) below.

\begin{figure}[htb]
\begin{center}
\epsfig{figure=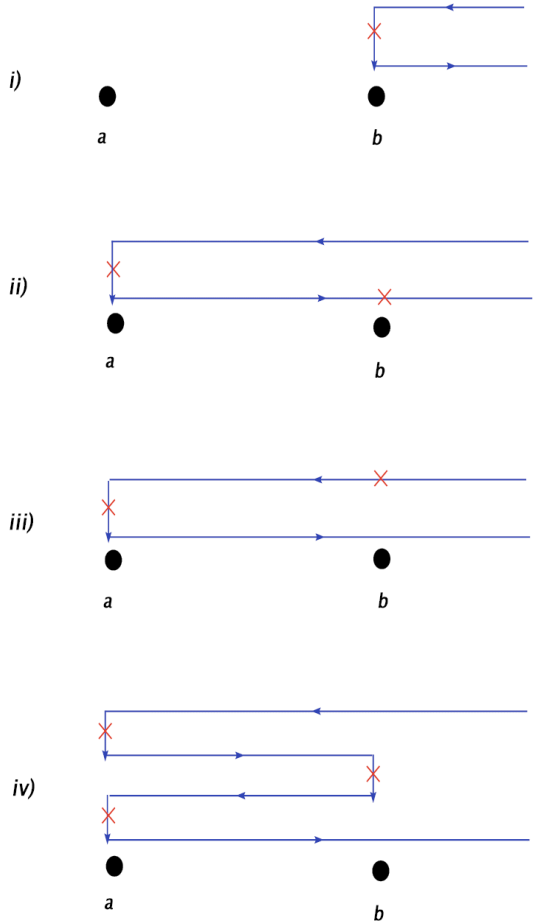,width=8.5cm}
\end{center}
\vspace*{-.6cm}
\caption{(Color online) Different paths contributing to $r_{R,1}$ at first 
order in $b$; $a$ and $b$ denote the static and oscillating potentials 
respectively. The crosses denote the scattering events.}
\end{figure}

We now observe that, to second order in the amplitude $b$, Eqs. 
(\ref{ir1}-\ref{rt}) and Eqs. (\ref{alr}-\ref{irb}) for the 
pumped current remain 
valid for the case of a general static potential which may be extended over 
more than one site, as long as the oscillating potential lies outside the
region of the static potential; we only have to use the appropriate 
expressions for the reflection and transmission amplitudes $r_{R,\pm 1} (k)$
and $t_{R,\pm 1} (k)$, and the bound state wave function $\psi_B (r)$. The
same derivation which was used to obtain the above equations for the case of 
a static potential at one site will work for a more general case. We can show
this in a different way by using the idea of sum over paths. To first order 
in $b$, the reflection amplitude $r_{R,1}$ in Eq. (\ref{rt}) can be understood
as arising from the sum of the following four paths (see Fig. 1), remembering 
that the oscillating potential lies to the right of the static potential. An 
electron incident from the far right with a wave number $k_0$ can be

\noi (i) reflected to the right by the oscillating potential to a wave 
number $k_1$ with an amplitude $-ib/(4 \ga \sin k_1)$ [this amplitude can be 
derived using the Born approximation on a lattice \ct{agarwal3} and 
remembering that $b \cos (\om t) = (b/2) (e^{i \om t} + e^{-i \om t})$], or

\noi (ii) reflected to the right by the static potential with an amplitude 
$r_R (k_0)$, and then transmitted to the right by the oscillating potential 
to a wave number $k_1$ with an amplitude $-ib/(4 \ga \sin k_1)$, or

\noi (iii) transmitted to the left by the oscillating potential to a wave 
number $k_1$ with an amplitude $-ib/(4 \ga \sin k_1)$, and then reflected to 
the right with an amplitude $r_R (k_1)$, or

\noi (iv) reflected to the right by the static potential with an amplitude 
$r_R (k_0)$, then reflected to the left by the oscillating potential to a wave
number $k_1$ with an amplitude $-ib/(4 \ga \sin k_1)$, and finally reflected 
to the right by the static potential with an amplitude $r_R (k_1)$.

\noi Similarly, the transmission amplitude $t_{R,1}$ in Eq. (\ref{rt}) can 
be understood as a sum of the following two paths. An electron incident 
from the far left with a wave number $k_0$ can be

\noi (i) transmitted to the right by the static potential with an amplitude
$t_R (k_0)$, and then transmitted to the right by the oscillating potential 
to a wave number $k_1$ with an amplitude $-ib/(4 \ga \sin k_1)$, or

\noi (ii) transmitted to the right by the static potential with an amplitude
$t_R (k_0)$, then reflected to the left by the oscillating potential to a wave
number $k_1$ with an amplitude $-ib/(4 \ga \sin k_1)$, and finally reflected 
to the right by the static potential with an amplitude $r_R (k_1)$.

Similar ideas can be used to derive the expressions in Eq. (\ref{alr}).
The amplitude $\al_R$ is the sum of two terms. An electron with a wave
function $\psi_B (r)$ can either be (i) transmitted to the right by the 
oscillating potential to a wave number $k_1$ with an amplitude $-ib/(4 
\ga \sin k_1)$, or (ii) transmitted to the left by the oscillating potential
to a wave number $k_1$ with an amplitude $-ib/(4 \ga \sin k_1)$, and then
reflected to the right by the static potential with an amplitude $r_R (k_1)$.
The amplitude $\al_L$ corresponds to an electron with a wave function 
$\psi_B (r)$ being transmitted to the left by the oscillating potential
to a wave number $k_1$ with an amplitude $-ib/(4 \ga \sin k_1)$, and then
transmitted to the left by the static potential with an amplitude $t_L (k_1)$.

We thus see that Eqs. (\ref{rt}) and (\ref{alr}) are valid to first order
in $b$ for any static potential, provided that the static and oscillating
potentials are separated by a finite distance.

\section{Equation of motion method}

We will now discuss how the pumped current can be obtained by numerically 
solving the equation of motion for the density matrix of a system with a 
finite number of sites \ct{agarwal1}. The density matrix of the system 
evolves according to the equation of motion
\beq \hat{\rho}(t+dt) ~=~ e^{-i\hat{H} (t) dt} ~\hat{\rho}(t) ~e^{i\hat{H} 
(t) dt} ~, \label{rhot} \eeq
where $\hat{H} (t) = {\hat H}_0 + {\hat V} (t)$ is given in Eqs. (\ref{h0}) 
and (\ref{pot}). The current across any bond is then obtained by taking
the trace of the appropriate current operator with $\hat{\rho}$. The current
operator on the bond from site $n$ to site $n+1$ and its expectation value
at time $t$ are given by
\bea \hat{j}_{n+1/2} &=& iq \ga ~(c_{n+1}^\da c_n -c_n^\da c_{n+1})~, \non \\
{\rm and} \quad j_{n+1/2} (t) &=& Tr (~\hat{\rho} (t) ~\hat{j}_{n+1/2} ~)
\non \\
&=& iq \ga ~[ \hat{\rho}_{n,n+1}(t) -\hat{\rho}_{n+1,n}(t) ]. \label{jnt} \eea
The charge transferred between the left and right halves of the system $L$ 
and $R$ between two times can be found either by integrating the above 
expression in time, or by taking the operator
\beq \De {\hat Q} ~=~ \frac{q}{2} ~\bigl[ ~\sum_{n \in R} ~c_n^\da c_n ~-~
\sum_{n \in L}~ c_n^\da c_n ~\bigr] ~, \label{dqt} \eeq
and computing $Tr~ (\hat{\rho} (t) \De \hat{Q})$ at the two times; these
methods give the same result for the charge transferred in a cycle.

In all our calculations, we take the left and right leads to have $N_l$ sites
each and the wire in the middle to have $N_w$ sites; the total number of sites
is $N=2N_l +N_w$. We take the density matrix at time $t=0$ to be given by that
of a single system governed by the Hamiltonian $H_0$ in Eq. (\ref{h0}) with 
$N$ sites, chemical potential $\mu$, and temperature $T$. If $E_\al$ and 
$\psi_\al (n)$ are the eigenvalues and eigenstates of the ${\hat H}_0$ ($\al$ 
and $n$ label the states and sites respectively), the initial density matrix 
is given by
\beq {\hat \rho}_{mn} (0) ~=~\sum_\al ~\psi_\al (m) ~\psi_\al^* (n) ~f(E_\al).
\label{rho0} \eeq
We then evolve the density matrix in time and compute the current and charge
transferred using Eqs. (\ref{rhot}-\ref{dqt}). 

An important point to note is that the finite length of the leads (with
$N_l$ sites) implies that the system has a return time equal to $2N_l /
v_F$ where the Fermi velocity $v_F = 2 \ga \sin k_F$ \ct{dhar}; this is the
time required for an electron to travel from the wire in the middle to the end
of either of the two leads and then return to the wire. The numerical results
can be trusted only for times which are less than the return time. Further, 
there are transient effects which last for one or two cycles; the effects of 
different choices of the initial density matrix get washed out after this 
transient period. All the numerical results presented below are therefore 
taken from times which are larger than the transient time but smaller than 
the return time; typically, we have computed the charge transferred between 
the times $6\pi/\om$ and $10\pi/\om$, where $\om$ is the oscillation
frequency. The dc part of the charge pumped per cycle should of course 
be independent of the location of the bond where it is measured; we have 
checked that this is true for all our numerical results except for the 
contribution of a bound state as we will discuss below.

\begin{figure}[htb]
\begin{center}
\epsfig{figure=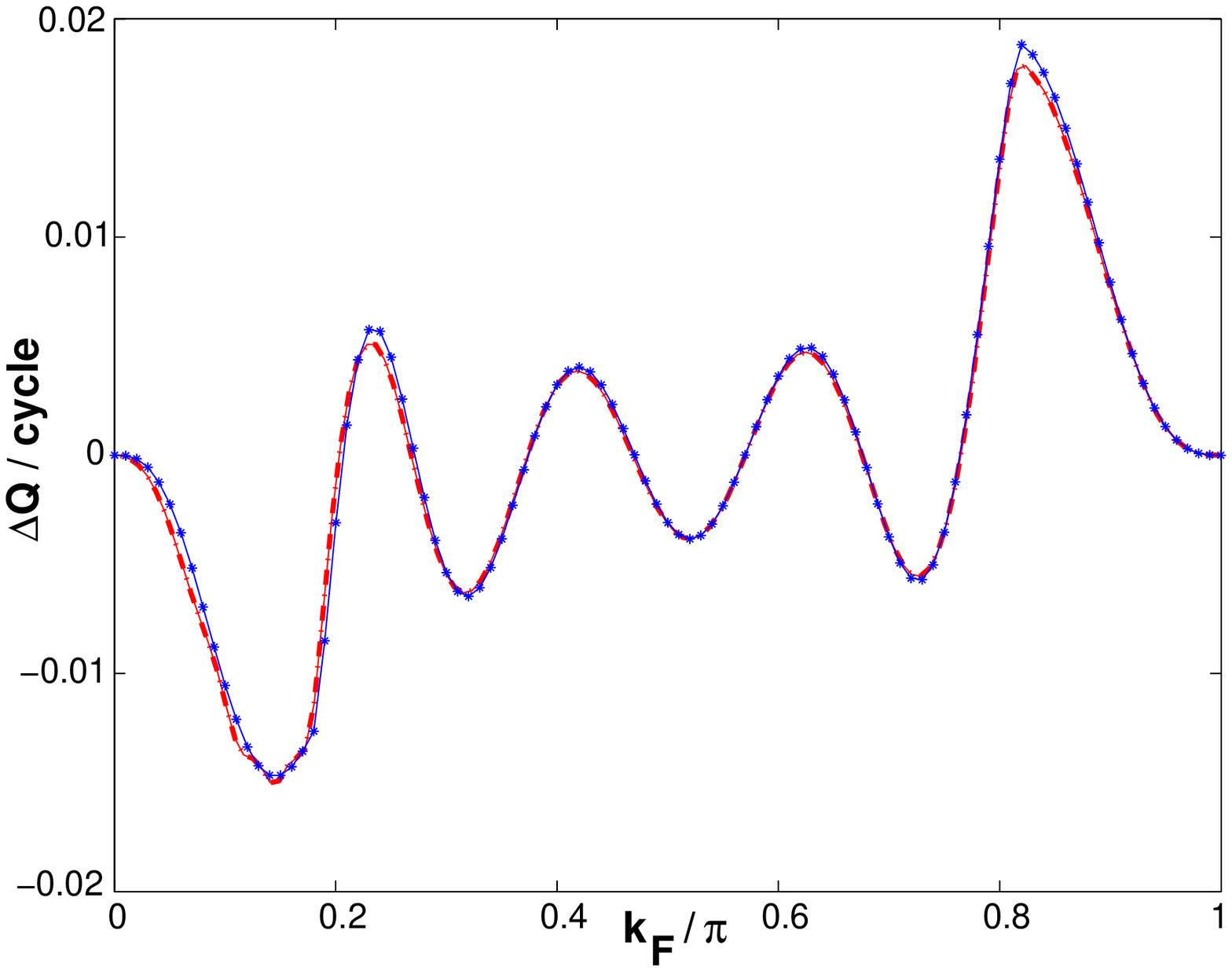,width=8.6cm}
\end{center}
\caption{(Color online) Charge pumped per cycle from left 
to right versus the Fermi wave number for a static potential separated from 
an oscillating potential by five sites, with $a=1.5$, $b = 0.25$, and $\om = 
\pi /10$; the system has 252 sites. The numerical and analytical results are 
shown by dash-dot (red) and starred (blue) lines, respectively.}
\end{figure}

\begin{figure}[htb]
\begin{center}
\epsfig{figure=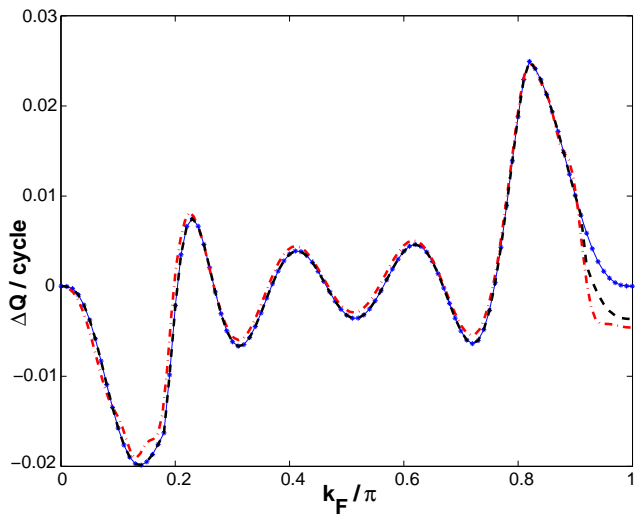,width=8.6cm}
\end{center}
\caption{(Color online) Charge pumped per cycle from left to right versus the 
Fermi wave number for a static potential separated from an oscillating 
potential by five sites, with $a=1$, $b = 0.25$, and $\om = \pi /10$; the 
system has 252 sites. The numerical and analytical results with and without 
the bound state contribution are shown by dash-dot (red), starred (blue) and 
dashed (black) lines, respectively.}
\end{figure}

We will now present our numerical results for the pumped charge (in units of
$q$). In all cases we have set the temperature to zero, the hopping amplitude 
$\ga =1$, and $r=5$, i.e., the static and oscillating potentials 
are separated by five lattice spacings. 

In Fig. 2, we show the charge pumped per 
cycle ($\De Q = (2\pi /\om) I_R$) as a function of the Fermi wave number
$k_F$ for the case $a=1.5$, $b = 0.25$, and $\om = \pi /10$ (corresponding 
to a time period of 20). The dash-dot and starred lines show the numerical 
and analytical results obtained from Eq. (\ref{ir1}) respectively; the 
agreement between the two is excellent. The pumped charge can be seen to go 
to zero at the band edges as expected. Although there is a positive energy 
bound state at $E_B = 2.5$, it does not contribute to the pumped charge 
because the first side band lies at an energy of $E_{-1} = E_B - \om \simeq 
2.19$ which is above the continuum band.

Figure 3 shows the charge pumped per cycle as a function of $k_F$ for $a=1$,
$b = 0.25$, and $\om = \pi /10$. The dash-dot line shows the numerical 
results, while the starred and dash lines show the analytical results obtained
from Eq. (\ref{ir1}) (continuum states) and Eq. (\ref{ir1}) plus (\ref{irb}) 
(continuum and bound states) respectively. In contrast to the case shown in 
Fig. 1, the total pumped current does not go to zero near $k_F = \pi$. This is
because of the contribution from a positive energy bound state; this has the 
energy $E_B \simeq 2.24$, and the first side band lies at an energy of $E_{-1}
= E_B - \om \simeq 1.92$ which lies within the continuum band. Hence, when
the chemical potential exceeds $E_{-1}$, the bound state begins
to contribute to the pumped charge. The above value of $E_{-1}$ 
corresponds to a Fermi wave vector of $k_F = \cos^{-1} (-1.92/2)
= 2.86$; we can see from the figure that when $k_F/\pi$ exceeds $2.86/\pi
\simeq 0.91$, the total pumped 
charge begins to deviate from the continuum contribution (which, according to 
Eq. (\ref{ir1}), does go to zero as $k_F \to \pi$). Using Eq. (\ref{irb}), we 
can compute the contribution to the pumped charge from the bound state, 
$\De Q_B = (2 \pi /\om) I_{RB}$. We find that $\De Q_B = - 0.0055$ which
agrees reasonably well with the value obtained numerically when $k_F \to \pi$.
However, we find that the pumped charge arising from the bound state has 
rather long lived transients, and the value of the dc part of the current 
varies significantly depending on the location of the bond where it is 
measured; the numerical result shown in Fig. 3 is the current measured at the 
bond lying midway between the static and oscillating potentials. Thus our 
model has a numerical difficulty in correctly computing the contribution of a
bound state to the current. The reason for this difficulty will be discussed 
in Sec. IV.

Figure 4 shows the charge pumped per cycle as a function of the oscillation
frequency $\om$ for $a=1.5$, $b = 0.25$, and $k_F = \pi /4$ (corresponding
to $\mu \simeq -1.414$). The dash-dot
and starred lines show the numerical and analytical results obtained from Eq. 
(\ref{ir1}) respectively. A noticeable change is seen to occur when $\om$
crosses a value of about $0.59$; this is because the first side band $E_{-1} 
= \mu - \om$ goes below the continuum band and stops contributing to the 
pumped charge at that point. A less noticeable change occurs for a similar
reason when $\om$ crosses a value of about $3.41$, where the first side band
$E_1 = \mu + \om$ goes above the continuum band. We also note that the pumped
charge goes to zero at $\om = 4$, as we had commented after Eq. (\ref{ir1}).

\begin{figure}[htb]
\begin{center}
\epsfig{figure=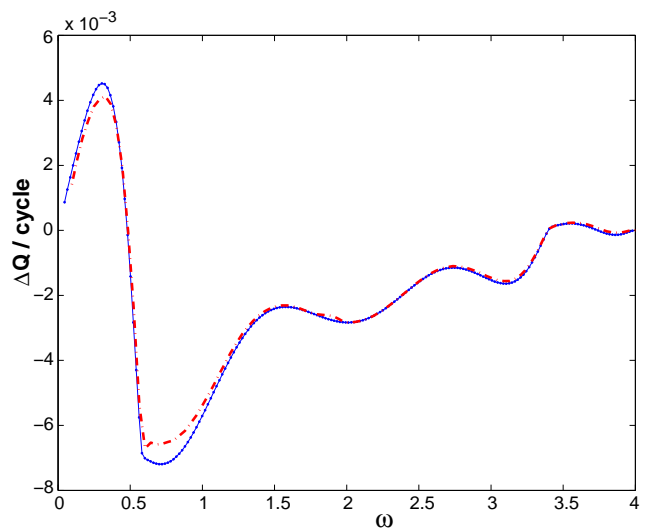,width=8.6cm}
\end{center}
\caption{(Color online) Charge pumped per cycle from left to right versus the 
oscillation frequency for a static potential separated from the oscillating 
potential by five sites, with $a=1.5$, $b = 0.25$, and $k_F = \pi /4$; the 
system has 308 sites. The numerical and analytical results are shown by 
dash-dot (red) and starred (blue) lines, respectively.}
\end{figure}

Finally, Fig. 5 shows the charge pumped per cycle as a function of the 
oscillation amplitude $b$ for $a=1.5$, $\om = \pi /10$, and $k_F = \pi /4$.
In this case, we cannot use the analytical expression given in Eq. (\ref{ir1})
since $b$ is not small in general and there is a substantial contribution from
higher side bands with $|p| \ge 2$. It is interesting to note that the pumped
charge vanishes and changes sign at certain values of $b$.

\begin{figure}[htb]
\begin{center}
\epsfig{figure=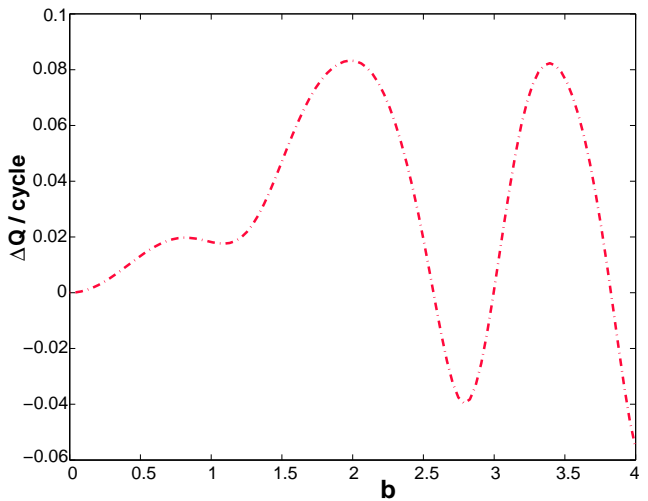,width=8.6cm}
\end{center}
\caption{(Color online) Charge pumped per cycle from left to right versus the 
oscillation amplitude for a static potential separated from the oscillating 
potential by five sites, with $a=1.5$, $\omega = \pi /10$, and $k_F = \pi /4$;
the system has 252 sites.}
\end{figure}

\section{Discussion}

In this paper, we have studied charge pumping by a single oscillating 
potential when the spatial left-right symmetry is broken due to the presence 
of a static potential at another point; this model is similar to the one used
to describe some recent experiments \ct{kaestner}. We have used Floquet
scattering theory to compute the pumped charge to second order in the 
oscillation amplitude. We have shown that if the oscillation frequency is 
larger than a threshold value, a bound state can contribute to charge 
pumping; this possibility does not seem to have been studied earlier.

For small amplitudes, we find that the results obtained numerically using the
equation of motion method generally agree well with the analytical results.
However, some numerical problems arise when the contribution from a bound 
state becomes important. These problems have been observed earlier in Ref. 
\ct{dhar} (see also Ref. \ct{stefan}); they are due to a difficulty in 
maintaining the occupation of the bound state at the correct equilibrium 
value. The simple model we have used for the numerical calculations has no 
interactions between electrons and no phonons which can lead to 
inelastic scattering processes and thereby maintain the occupation of the 
bound state at a value dictated by the Fermi function. Some ways of addressing
the problem of bound states have been discussed in Refs. \ct{dhar,stefan}.

It would be interesting to generalize our analysis to the case of an arbitrary
time-dependent potential applied at one point where the potential may contain
a very large number of oscillation frequencies with arbitrary amplitudes. In 
particular, one can consider the case of a noisy potential and study whether 
that can, on the average, pump charge in one direction when the left-right 
symmetry is broken by a static potential. The brief discussion in Sec. II of 
a potential consisting of a few oscillating terms with low frequencies and 
small amplitudes suggests that a weak noise may indeed be able to pump charge
on the average, but a detailed investigation of this may be useful.

\section*{Acknowledgments}

A.A. thanks CSIR, India for financial support. D.S. thanks A. Dhar for many
stimulating discussions. We thank DST, India for financial support under 
the projects SR/FST/PSI-022/2000 and SR/S2/CMP-27/2006.

\end{document}